# Tailored $Zn_{(1-x)}Ti_xAl_2O_4$ Nanocomposite Particles via Sol-Gel Route for High-Performance Humidity Sensing

Ramalakshmi K
*Department of Electronics and Communications Engineering, GIET University, Odisha, India*

Sasmita Dash
*Department of Electronics and Communications Engineering, GIET University, Odisha, India*

Srilali Siragam
*Department of Electronics and Communications Engineering, Swarnandhra College of Engg. & Tech.,* Andhra Pradesh, India

*Abstract*— **Humidity sensors play a vital role in industrial, healthcare, agricultural, and environmental applications; however, conventional sensors often suffer from issues like low sensitivity, slow response, and poor stability. This study investigates sol-gel synthesized $Zn_{0.85}Ti_{0.15}Al_2O_4$ nanocomposite ceramics for high-performance humidity sensing. X-ray Diffraction (XRD) analysis confirms a nanocrystalline structure, while Transmission Electron Microscopy (TEM) reveals nearly spherical grains with sizes ranging from 11 to 14 nm. The optical bandgap of 3.76 eV indicates the enhanced sensing potential. The sensor exhibits a significant decrease in resistance, from 500 MΩ (15% RH) to 90 MΩ (90% RH), with fast response (50 s) and recovery times (50 s). Low Hysteresis values (5.86% at 30% RH, 7.69% at 60% RH, and 4.28% at 90% RH) highlight high sensitivity, stability, and repeatability. These results indicate the potential of $Zn_{0.85}Ti_{0.15}Al_2O_4$ as a promising material for next-generation resistive humidity sensors suitable for commercial and industrial deployment.**

## I. INTRODUCTION

Nanostructures are materials with at least one dimension ranging from 1 to 100 nm, exhibiting unique physicochemical and plasmonic properties that drive applications from research to commercialization. Nanostructure offers higher melting temperatures, enhanced conductivity, superior light absorption, optical sensitivity, catalytic activity, and improved wettability than bulk materials. These properties make them essential for energy efficiency, structural reinforcement, antibacterial surfaces, self-cleaning materials, and sensors.

Gas and humidity sensors based on nanotechnology are vital for environmental monitoring, healthcare, agriculture, and industry, whereas traditional sensors suffer from large size, low sensitivity and slow response. In contrast, nanomaterials significantly enhance sensor performance due to their unique properties such as quantum effect, reactivity, and high surface area-to-volume ratio [1]. Specifically, the nanomaterial graphene gained significant interest in sensing and electromagnetic applications due to its unique band structure and material properties [2]. Graphene nanomaterial-based devices perform better than their counterparts [3]. Graphene is highly sensitive to ammonia ($NH_3$), nitrogen dioxide ($NO_2$), and volatile organic compounds (VOCs), while carbon nanotubes (CNTs) detect ammonia ($NH_3$), oxygen($O_2$), and hydrogen sulphide ($H_2S$). Conductive polymers like polyaniline (PANI), polypyrrole (PPy), and polythiophene (PTh) respond to humidity and gases such as carbon dioxide ($CO_2$) and ammonia ($NH_3$). Hybrid materials, such as polymer-metal oxide composites, further improve sensitivity and stability. Hisashi.S et al. demonstrated that metals such as platinum (Pt), silver (g) and gold (Au) act as catalysts, accelerating gas adsorption and reaction, particularly for Carbon Monoxide (CO) and Hydrogen ($H_2$). Particularly for resistive gas sensors, nanomaterial-based metal oxides, such as titanium oxide ($TiO_2$), zinc oxide (ZNO), and tin oxide ($SnO_2$), are widely used for their stability and affordability [4].

R. Sakthivel et al. demonstrated enhanced Nitrogen Dioxide ($NO_2$) gas sensing using graphene-decorated Nickel disulfide ($NiS_2$) thin films [5]. Platinum single atoms (Pt SAs) and tin oxide nanorods ($SnO_2$ NRs) are sequentially integrated onto Silicon Carbide (SiC) nanosheets to leverage their catalytic effects. Rani et al. fabricated SnSe nanostructured films using in-house synthesized SnSe powder via thermal evaporation, demonstrating industrial viability. The films exhibited strong $NO_2$ sensing at room temperature [6]. In another study, Kawther Assili et al. synthesized SnSe films via CVD using an organophosphorus selenium precursor at temperatures ranging from 200–500 °C. SnSe films deposited on alumina substrates exhibited excellent methane-sensing properties, with higher sensitivity observed at 200 °C compared to $SnSe_2$ [7].

Parthibavarman et al. synthesized SnO hexagonal nanodiscs and $SnO_2$ spherical nanoparticles via a microwave irradiation method. The $SnO_2$ nanoparticles exhibited excellent humidity-sensing performance [8]. The improved performance is linked to increase in oxygen vacancies and enhanced crystallinity. Zaihua Duan et al. fabricated a high-performance humidity sensor using halloysite nanotubes (HNTs), leveraging their hydrophilic nature, hollow structure, and large surface area [9]. H. Dai et al. developed a composite humidity sensor composed of chitosan (CS), single-walled carbon nanotubes and ZnO. These sensors hold potential for applications in textiles, cosmetics, electronics, and protective coatings [10].

These advancements in gas and humidity sensing demonstrate the effectiveness of nanostructured materials and heterojunction engineering in enhancing sensitivity, response speed, and selectivity. Materials like graphene, SiC, SnSe, SnO2, HNTs, and chitosan-based composites are paving the way for the future generations with low-power, high-performance, and scalable sensors for industrial, healthcare, and environmental applications.

As we know, nanomaterials have high surface-to-volume ratio than bulk materials. Many researchers can develop sensors with individual nanomaterials or nanocomposite materials, and only few can develop the nanocomposite ceramics with higher thermal and chemical stability than the nanocomposites in sensing applications. This study mainly focused on the sol-gel synthesis of $Zn_{(1-x)}Ti_xAl_2O_4$ nanocomposite ceramics and their different characterizations, such as structural, optical, morphological, and chemical characterization. These nanoparticles were utilized to develop humidity sensors, and their sensing performance was evaluated. The primary objective is to modify the properties of $ZnAl_2O_4$ by incorporating $TiO_2$, making it suitable for humidity sensing applications. To the best of our knowledge, $Zn_{(1-x)}Ti_xAl_2O_4$ composites have not been previously explored for this purpose, and Ti with 0.15 mg concentration explores better sensing results than the remaining concentrations. Section 2 details the synthesis process, Section 3 presents the results, and Section 4 concludes the study.

## II. SYNTHHESIS OF $ZN_{(1-x)}TI_xAL_2O_4$

The sol-gel process is used to synthesize nanoparticles from aluminium nitrate nonahydrate ($Al_2(NO_3)_3.9H_2O$, Sigma Aldrich), titanium-iso-peroxide (TTIP), zinc acetate ($CH_3COO)_2Zn2H_2O$ (LR-grade), ethanol ($C_2H_5OH$), ethylene glycol (EG), distilled water, and nitric acid ($HNO_3$) without purification.

A beaker with 45 ml of ethanol is heated to 75 °C by including 3.5 ml of distilled water and agitating it at 350 rpm with a magnetic stirrer. Upon reaching this temperature, 2.5 ml of TTIP is introduced, and the mixture is subsequently dried at 70 °C after being maintained at 75 °C for three hours until gelation occurs. Independently, 10 ml of ethanol is employed to dissolve 2.1 g of $Al_2(NO_3)_3.9H_2O$. The mixture is stirred for 6-10 minutes, after which 0.1 ml of ethylene glycol, 0.25 g of titanium oxide, and 1.08 g of zinc acetate are incorporated. The mixture is heated to 85 °C for one hour, placed in a hot oven at 180 °C for half an hour, chilled for nexttt half an hour, and then calcined for ninety minutes at 700 °C in a muffle furnace. A uniform consistency is attained by milling the resultant white powder. The synthesized nanoparticles were characterized for crystallinity and phase using X-ray diffraction (XRD), structural analysis through Raman spectroscopy, morphological investigation via transmission electron microscopy (TEM), and stoichiometry assessment using energy-dispersive X-ray spectroscopy (EDS) integrated with TEM. Additionally, ultraviolet-visible spectroscopy (UV–Vis) was employed to define the optical band gap of the prepared material.

## III. RESULTS AND DISCUSSION

XRD analysis of the ZnTiAl2O4 sample reveals crystallinity around 20-80°, with anatase (a) and rutile (r) TiO2 present with ZnO. A wurtzite ZnO peak was found at 2θ = 56.64° and 62.25° (planes 110 and 103). ZnAl2O4 peaks on planes 220, 311, 400, 331, 422, 511, 440, 620, and 533 occur at 2θ = 31.4°, 36.9°, 44.8°, 49.19°, 55.64°, 59.5°, 65.25°, 73.98°, and 77. At 25.4° (101), a-TiO2 is found, whereas r-TiO2 is at 27.5° (110). By Srilali S et al., if concentration of TiO2 increases, the full width at half-maxima of ZnAl2O4 diffraction peaks decreases [11]. By using Scherrer's formula for crystal size calculation,

$$D_{hkl} = \frac{k\lambda}{\beta \cos\theta} \qquad \text{------- (1)}$$

Where k is the Scherrer constant, which is 0.94 for spherical-shaped crystals.

The crystallite size of $Zn_{0.85}Ti_{0.15}Al_2O_4$ sample was estimated to be 9.7 nm, corresponding to the plane (220) of ZnAl2O4. The findings match JCPDS files 80-0074, 05-0669, 21-1272, 21-1276 and previous reports [12]. The concentration of TiO2 considerably affects the size and crystallinity of the composite.

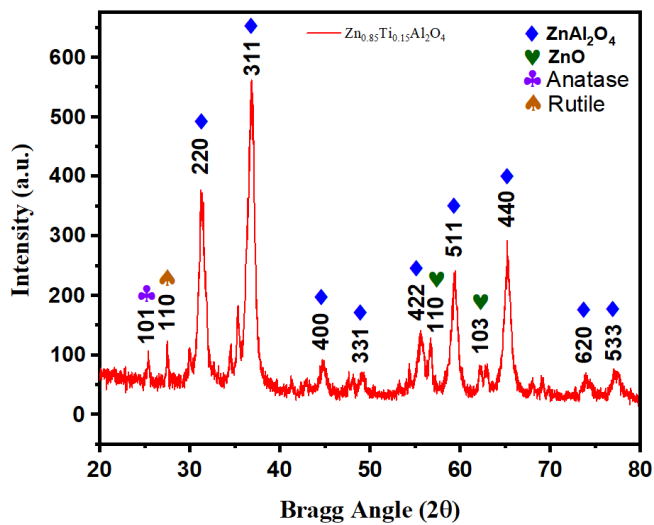

Fig. 1 XRD response of the prepared $Zn_{0.85}Ti_{0.15}Al_2O_4$ sample

Figure 2 illustrates the Raman spectra of $Zn_{(1-x)}Ti_xAl_2O_4$ nanocomposite ceramics with varying TiO2 concentration ($x$ = 0.15 mol). The peak at 437 cm⁻¹ corresponds to the E2 mode, characteristic of the active optical phonon mode in the ZnO wurtzite crystalline structure. Another peak at 268 cm⁻¹ is associated with the B1 mode, attributed to structural defects. Additionally, a peak at 632 cm⁻¹ is identified as the active mode of ZnAl2O4. The $Zn_{0.85}Ti_{0.15}Al_2O_4$ sample exhibits anatase- and rutile-TiO2 phases as TiO2 concentration increases. Anatase-TiO2 peaks at 391, 513, and 632 cm⁻¹, while rutile-TiO2 peaks at 237 and 437 cm⁻¹. These observations align well with previously reported in Nasr M et al., literature [13].

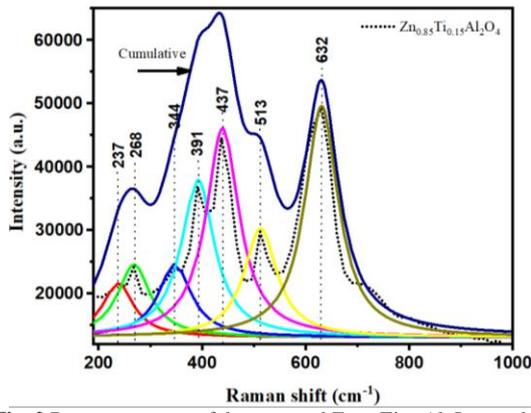

**Fig. 2** Raman response of the prepared $Zn_{0.85}Ti_{0.15}Al_2O_4$ sample

With increased $TiO_2$ concentration, the appearance of anatase- and rutile-$TiO_2$ peaks is evident alongside the prominent characteristic peaks of the composite. We have also identified a small peak at 344 $cm^{-1}$.

TEM is used to analyze $Zn_{0.85}Ti_{0.15}Al_2O_4$ nanoparticle morphology. TEM and EDS spectra are shown in Figure 3. A TEM image of the produced sample showed clear boundaries of spherical grains in Fig. 3a. Grain distribution is shown in an inset histogram. Approximately 14 nm was the granular diameter. EDS analysis of $Zn_{0.85}Ti_{0.15}Al_2O_4$ shows its elemental composition in Fig. 3b and confirms Zn (8.62, 9.58 eV), Al (1.48), O (0.525), and Ti (4.93 eV).

The bandgap influences humidity sensor performance by affecting electrical and optical properties. Water adsorption alters the band structure, impacting conductivity and carrier concentration. In semiconductors like ZnO (bandgap 3.27 eV direct, 2.96 eV indirect) and $TiO_2$, adsorption enhances ionic conductivity. $Zn_{0.85}Ti_{0.15}Al_2O_4$ nanostructures with a 3.76 eV bandgap improve sensitivity, enabling efficient water molecule adsorption and desorption for enhanced humidity sensing. Figure 4 shows the optical band gap of $Zn_{(1-x)}Ti_xAl_2O_4$ nanoparticles after the annealing treatment at 700 °C for one hour. The method developed by Wood and Tauc was used to estimate the optical band gap energy ($E_g$).

$$\alpha = \frac{1}{h\upsilon}[A(h\upsilon - E_g)]^{1/2} \quad \text{(Direct)} \text{-----------} \quad (2)$$

$$\alpha = \frac{1}{h\upsilon}[A(h\upsilon - E_g)]^2 \quad \text{(Indirect)} \text{----------} \quad (3)$$

where $\alpha$ is the coefficient of absorption, $h\upsilon$ is the energy of the photon, $E_g$ is the band gap energy, and A is a constant depending on the type of transition. There is no single formula for defining the relation between band gap energy and adsorption. The Tauc equation defines how the band gap influences the material's absorption property, which plays a key role in defining the adsorption property.

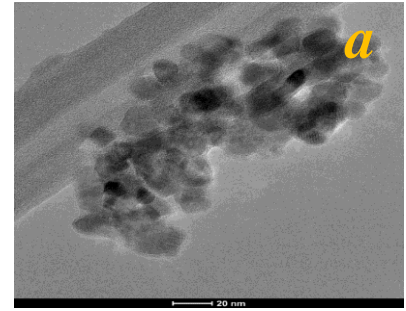

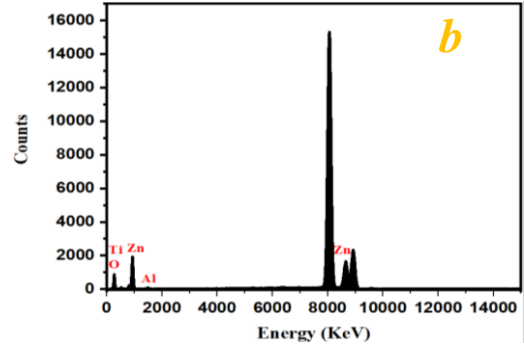

**Fig. 3** (a) TEM (b) EDS response of the prepared $Zn_{0.85}Ti_{0.15}Al_2O_4$ sample

Our material reported the $E_g$ of $ZnAl_2O_4$-doped $TiO_2$ is ~3.29 eV for the direct gap and also the indirect gap is 2.96 eV. In addition, by increasing intermediary energy levels within the optical band gap, Eg will be decreased. S. Dash et al., demonstrated that the decreased band gap shows that the Eg can improve optical characteristics, making it useful for sensing devices [14].

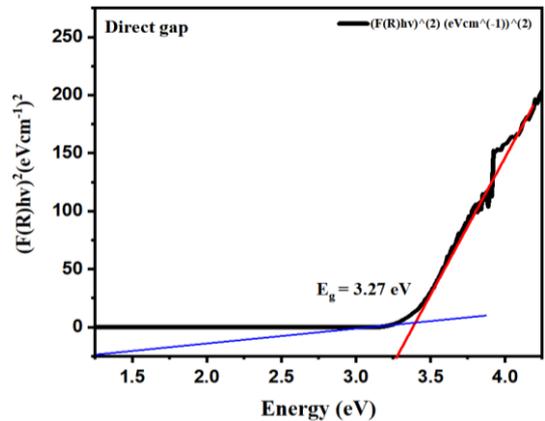

**Fig. 4** Optical band gap (Direct) of $Zn_{0.85}Ti_{0.15}Al_2O_4$ sample

Adsorption and desorption are crucial when water vapour molecules interact with humidity sensors. Physical or chemical adsorption of environmental water molecules occurs on the sensor material. Most humidity sensors use physisorption, where weak van der Waals forces help. Water molecules adsorb onto sensing materials, change their characteristics. Ionic conduction is increased by adsorbed water in resistive sensors, lowering resistance. In capacitive sensors, water adsorbed increases the dielectric constant and capacitance. The added mass from adsorbed water shifts the resonance frequency for mass-based sensors, while optical sensors detect refractive index or other optical qualities.

During desorption, water molecules leave the sensor and return to the environment. Reduced relative humidity reduces water molecule interaction. Dynamic equilibrium between adsorption and desorption controls sensor response time and accuracy. Adsorption and desorption speed up the reaction and recovery, making them essential for real-time humidity monitoring. Adsorption and desorption effectiveness vary with sensing material porosity, surface area, hydrophilicity, temperature, and humidity.

The resistance humidity hysteresis for $Zn_{0.85}Ti_{0.15}Al_2O_4$ is shown in Figures 5 & 6. Desorption is weaker than adsorption, as shown in the figure. This shows exothermic and endothermic reactions, while adsorption and desorption occurred at various rates. Desorption has slightly lower resistance than adsorption, reducing sample hysteresis. Less hysteresis is better for sensor development.

Figure 5 illustrates the relationship between electrical resistance and relative humidity for the $Zn_{0.85}Ti_{0.15}Al_2O_4$ sample during the adsorption process. The resistance is highest at ~500 MΩ when the relative humidity is 15% RH and gradually decreases as the humidity increases.

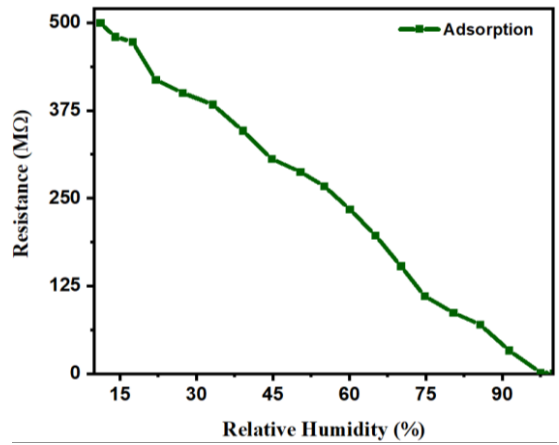

**Fig. 5** Resistance humidity sensor (Adsorption) hysteresis curve for $Zn_{0.85}Ti_{0.15}Al_2O_4$ sample.

At 30% RH, the resistance drops to 400 MΩ, followed by 320 MΩ at 45% RH and 220 MΩ at 60% RH. As humidity reaches 75% RH, the resistance further declines to 90 MΩ, and at 90% RH, it approaches 0 MΩ, indicating a strong response to humidity changes. This trend is attributed to the progressive adsorption of water molecules on the material's surface. Initially, a thin water layer forms, having minimal impact on conductivity. However, Ri Gum-Chol et al. demonstrated that as humidity rises, additional water molecules facilitate ionization ($H_2O \rightarrow H^+ + OH^-$), enhancing proton conduction and drastically reducing resistance [15]. Incorporating Ti in $ZnAl_2O_4$ may contribute to forming oxygen vacancies, improving water adsorption and overall sensitivity. These characteristics make $Zn_{0.85}Ti_{0.15}Al_2O_4$ a promising candidate for resistive humidity sensors due to its high sensitivity, broad detection range, and stable performance.

Humidity sensor sensitivity declines over time due to environmental factors, material degradation, and ageing.

New sensors are highly responsive due to clean, active adsorption sites, but surface contamination, swelling, or prolonged humidity exposure can reduce performance. Saturation effects and chemical interactions further degrade sensitivity. Nanostructured and composite sensors offer better longevity, while protective coatings and self-healing materials help mitigate degradation. Sensitivity, measured as signal change per RH variation, typically stabilizes initially, then declines. Using stable materials, encapsulation, and recalibration ensures long-term reliability and accuracy.

As humidity decreases, water molecules start desorbing, leading to an increase in resistance. At 75% RH, the resistance rises to approximately 90 MΩ, increasing to 220 MΩ at 60% RH. When RH drops to 45%, the resistance reaches 320 MΩ; at 30% RH, it increases to about 400 MΩ. Finally, at 15% RH, the resistance stabilizes at ~500 MΩ, indicating complete desorption. This behaviour is attributed to the removal of water molecules from the sensor's surface. Initially, weakly bound physisorbed water desorbs quickly, causing a gradual increase in resistance. As humidity continues to drop, chemisorbed water layers are removed, leading to a slower resistance increase. The effective desorption process suggests good reversibility and fast recovery time, confirming the material's high sensitivity, stability, and suitability for practical humidity sensing applications.

## IV. Conclusions

The sol-gel synthesized $Zn_{0.85}Ti_{0.15}Al_2O_4$ nanocomposite ceramics exhibited nanocrystalline structure (XRD) and spherical grains (~11–14 nm, TEM), with an optical bandgap of 3.76 eV. The sensor demonstrated a highly sensitive response to humidity, with resistance decreasing from 500 MΩ at 15% RH to 90 MΩ at 90% RH. The fast response (50 s) and recovery times (50 s), along with low hysteresis (5.86% at 30% RH, 7.69% at 60% RH, and 4.28% at 90% RH), confirm its stability and efficiency for humidity sensing. The presence of Ti in $ZnAl_2O_4$ enhances adsorption sites, improving sensor performance. These findings suggest that $Zn_{0.85}Ti_{0.15}Al_2O_4$ is a promising material for practical resistive humidity sensors, with potential for commercial and industrial applications. Future studies may explore cyclic stability tests, evaluation under varying temperatures, long-term durability and optimization for real-world integration.

### References

1.  Ramalakshmi K, Sasmita Dash, Srilali Siragam, "ZnAl$_2$O$_4$ Based Nanomaterial for Sensing Application", 3rd International Conference on Functional Materials for NexGen Applications (ICFMNA), Kalavakkam, Tamilnadu, February 03-04, 2025.
2.  S. Dash, C. Liaskos, I. F. Akyildiz and A. Pitsillides, "Graphene Hypersurface for Manipulation of THz Waves", Materials Science Forum, Vol. 1009, pp 63-68, 2020.

3. S. Dash, and A. Patnaik, "Performance of Graphene Plasmonic Antenna in Comparison with their Counterparts for Low-Terahertz Applications", Plasmonics, Vol. 13, no. 6, pp. 2353-2360, 2018.

4. Yasuhisa Naitoh, Touru Sumiya, Hisashi Shima, Hiroyuki Akinaga, "High-speed hydrogen sensor fabricated using a platinum/titanium oxide nanocontact", Sensors and Actuators B: Chemical, Vol. 371, pp 132531, 2022.

5. Sakthivel R., Geetha A., Anandh B.A. et al. "Thin films of graphene decorated with $NiS_2$ hybrid sensor for detection of $NO_2$ gas". J Mater Sci: Mater Electron, vol. 33, pp. 23404–23417, Sept.2022.

6. S. Rani, M. Kumar, H. Sheoran, R. Singh, V. Nand Singh, "Rapidly responding room temperature $NO_2$ gas sensor based on SnSe nanostructured film". J.Mater. Today Commun., Vol. 30, pp. 103-135, March.2022.

7. Kawther Assili, Oriol Gonzalez, Khaled Alouani, Xavier Vilanova "Structural, morphological, optical and sensing properties of SnSe and $SnSe_2$ thin films as a gas sensing material", Arabian Journal of Chemistry, vol. 13, pp.1229-1246, Jan.2020.

8. M. Parthibavarman, V. Hariharan, C. Sekar, "High-sensitivity humidity sensor based on SnO2 nanoparticles synthesized by microwave irradiation method", J.Materials Science and Engineering:C, vol. 31, pp.840-844, July.2011.

9. Zaihua Duan, Qiuni Zhao, Si Wang, Qi Huang, Zhen Yuan, Yajie Zhang, Yadong Jiang, Huiling Tai, "Halloysite nanotubes: Natural, environmental-friendly and low-cost nanomaterials for high-performance humidity sensor", J.Sensors and Actuators B: Chemical, vol. 317, pp.128204, Aug.2020.

10. Haipo Dai, Nana Feng, Jiwei Li, Jie Zhang, Wei Li, "Chemi resistive humidity sensor based on chitosan/zinc oxide/single-walled carbon nanotube composite film", J.Sensors and Actuators B: Chemical, vol. 283, pp.786-792 Sept.2019.

11. Srilali Siragam, R. S. Dubey, Pappula, Lakshman, "Synthesis & Study of Zinc Titanium Aluminate Nanoceramic Composite for Patch Antenna Application" 5th IEEE International WIE Conference on Electrical and Computer Engineering (WIECON-ECE), June 13, 2020.

12. Li, Z., Xing, L., Zhang, N., *et al.*: "Preparation and photocatalytic property of $TiO_2$ columnar nanostructure films", J.*Mater.Trans.*, vol. 52, pp.1939–1942 Apr.2011.

13. Nasr M, Viter R, Eid C, Warmont F, Habchi R, Mielea P, Bechelany M, "Synthesis of novel $ZnO/ZnAl_2O_4$ multi co-centric nanotubes and their long-term stability in photo catalytic application". J.RSC Adv, vol. 6, pp. 103692–103699, Oct.2016.

14. K. Ramalakshmi, Sasmita Dash, Srilali Siragam, "Next-generation sensors enabled by sol-gel derived $ZnO$–$ZnAl_2O_4$–$TiO_2$ nanoparticles", J.Materials Letters, Vol. 400, pp. 139121, Jul.2025.

15. Ri Gum-Chol, Kim Jin-Song and Yu Chol-Jun, "Role of Water Molecules in Enhancing the Proton Conductivity on Reduced Graphene Oxide under High Humidity" J. Phy.Rev.Appl., Vol. 10, pp- 034018, 2018.